\documentstyle{article}

\newcommand{\ba}{\begin{eqnarray}}
\newcommand{\ea}{\end{eqnarray}}
\input{psfig}
\def\ii{\'{\i}}
\begin{document}
\pagestyle{plain}

\title{Interacting Boson Models \\ 
of Nuclear and Nucleon Structure \\ \mbox{} \\ 
Modelos de Bosones Interactuantes \\ 
de la Estructura Nuclear y Nucle\'onica
\footnote{Pl\'atica plenaria invitada en el  
`XL Congreso Nacional de F\ii sica',
Monterrey, Nuevo Le\'on, 27 - 31 de octubre de 1997}}
\author{R. Bijker\\
Instituto de Ciencias Nucleares,\\ 
Universidad Nacional Aut\'onoma de M\'exico,\\
A.P. 70-543, 04510 M\'exico, D.F., M\'exico 
\and
A. Leviatan\\
Racah Institute of Physics, The Hebrew University,\\
Jerusalem 91904, Israel}
\date{}
\maketitle

\noindent
ABSTRACT: 
Interacting boson models provide an elegant and powerful method 
to describe collective excitations of complex systems by introducing 
a set of effective degrees of freedom. We review the interacting 
boson model of nuclear structure and discuss a recent extension 
to the nucleon and its excited states.\\
\mbox{}\\
RESUMEN: 
Los modelos de bosones interactuantes proveen un m\'etodo elegante y 
eficaz para describir excitaciones colectivas de sistemas complejos 
mediante la introducci\'on de grados de libertad efectivos. Primero 
revisamos el modelo de bosones interactuantes de la estructura nuclear 
y luego presentamos una extensi\'on a la estructura del nucle\'on y 
sus estados excitados.\\
\mbox{}

\clearpage

\section{Introduction} 

Atomic nuclei are complex systems consisting of large numbers of 
strongly interacting protons and neutrons and involving many degrees 
of freedom. In addition, the nucleons themselves are composite particles, 
each including three valence quarks. In principle, the structure and 
interactions of nucleons are described by quantum chromodynamics (QCD) 
of quarks and gluons which has emerged as the fundamental theory of 
strong interactions. However, the energy domain of nuclear physics, 
several MeV's ($=10^6$ eV's)
for nuclear structure and several GeV's ($=10^9$ eV's) for excitations 
of the nucleon, belongs to the nonpertubative regime of QCD for which,  
except for lattice calculations of ground state properties, 
no reasonable solution exists. 

Nevertheless, the low-lying spectrum of many nuclei exhibits a 
surprisingly simple structure. In the absence of an exactly solvable 
theory and reliable approximation methods, one has to rely on models 
of nuclear structure and symmetries to `understand' these regular features. 
In models one attempts to isolate the most important degrees of 
freedom and deal with them explicitly. Examples of nuclear models 
are the shell model in which the complicated motion of nucleons inside 
a nucleus is replaced by the motion of independent nucleons in a 
static spherical potential well \cite{Talmi}, the collective or 
geometric model in which collective nuclei are described in terms 
of geometric variables that characterize the shape and deformation 
of the nuclear surface \cite{BM}, and the interacting boson model 
in which collective quadrupole states in nuclei are described 
in terms of a system of interacting  
monopole and quadrupole bosons \cite{IBM}. 

Whereas in low energy nuclear physics it is a good approximation to 
neglect the internal structure of the nucleon, this is no longer the 
case for excitations of the nucleon itself (baryon resonances).   
Nowadays the nucleon is viewed as a confined system of quarks  
interacting via gluon exchange. Effective models of the nucleon 
are all based on three constituent parts that carry the internal degrees 
of freedom of spin, flavor and color \cite{eightfold}, but differ 
in their treatment of radial (or orbital) excitations. At the same time, 
the baryon mass spectrum shows some remarkable regularities, 
such as linear Regge trajectories and parity doublets, which indicates 
that a collective type of dynamics may play an important role in the 
structure of baryons.

In this contribution we show that interacting boson models provide an 
elegant and, at the same time, powerful method to describe collective 
excitations of complex systems by introducing a set of effective degrees 
of freedom. We first review the main features of the interacting boson 
model of nuclear structure, and next discuss a recent extension to 
the nucleon and its excited states. 

\section{Nuclear structure}

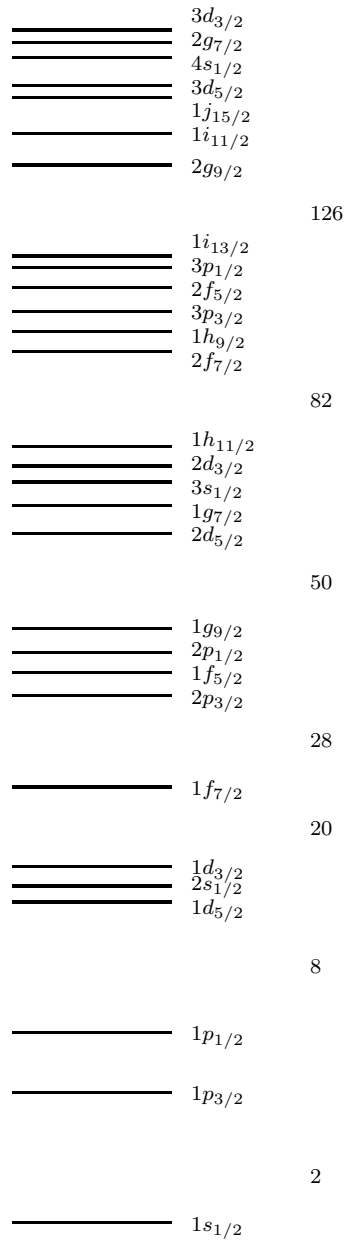
\begin{figure}
\centering
\setlength{\unitlength}{1.5pt}
\begin{picture}(100,320)(0,0)
\thicklines
\put ( 0,  0) {\line(1,0){40}}
\put ( 0, 33) {\line(1,0){40}}
\put ( 0, 48) {\line(1,0){40}}
\put ( 0, 81) {\line(1,0){40}}
\put ( 0, 85) {\line(1,0){40}}
\put ( 0, 90) {\line(1,0){40}}
\put ( 0,110) {\line(1,0){40}}
\put ( 0,133) {\line(1,0){40}}
\put ( 0,139) {\line(1,0){40}}
\put ( 0,144) {\line(1,0){40}}
\put ( 0,150) {\line(1,0){40}}
\put ( 0,174) {\line(1,0){40}}
\put ( 0,181) {\line(1,0){40}}
\put ( 0,187) {\line(1,0){40}}
\put ( 0,191) {\line(1,0){40}}
\put ( 0,196) {\line(1,0){40}}
\put ( 0,220) {\line(1,0){40}}
\put ( 0,225) {\line(1,0){40}}
\put ( 0,230) {\line(1,0){40}}
\put ( 0,236) {\line(1,0){40}}
\put ( 0,241) {\line(1,0){40}}
\put ( 0,244) {\line(1,0){40}}
\put ( 0,267) {\line(1,0){40}}
\put ( 0,275) {\line(1,0){40}}
\put ( 0,284) {\line(1,0){40}}
\put ( 0,287) {\line(1,0){40}}
\put ( 0,294) {\line(1,0){40}}
\put ( 0,298) {\line(1,0){40}}
\put ( 0,301) {\line(1,0){40}}
\thinlines
\footnotesize
\put (45, -2) {$1s_{1/2}$}
\put (75, 10) {2}
\put (45, 31) {$1p_{3/2}$}
\put (45, 46) {$1p_{1/2}$}
\put (75, 63) {8}
\put (45, 78) {$1d_{5/2}$}
\put (45, 84) {$2s_{1/2}$}
\put (45, 88) {$1d_{3/2}$}
\put (75, 98) {20}
\put (45,108) {$1f_{7/2}$}
\put (75,120) {28}
\put (45,131) {$2p_{3/2}$}
\put (45,137) {$1f_{5/2}$}
\put (45,143) {$2p_{1/2}$}
\put (45,149) {$1g_{9/2}$}
\put (75,160) {50}
\put (45,172) {$2d_{5/2}$}
\put (45,178) {$1g_{7/2}$}
\put (45,184) {$3s_{1/2}$}
\put (45,190) {$2d_{3/2}$}
\put (45,196) {$1h_{11/2}$}
\put (75,206) {82}
\put (45,216) {$2f_{7/2}$}
\put (45,222) {$1h_{9/2}$}
\put (45,228) {$3p_{3/2}$}
\put (45,234) {$2f_{5/2}$}
\put (45,240) {$3p_{1/2}$}
\put (45,246) {$1i_{13/2}$}
\put (75,253) {126}
\put (45,265) {$2g_{9/2}$}
\put (45,273) {$1i_{11/2}$}
\put (45,279) {$1j_{15/2}$}
\put (45,285) {$3d_{5/2}$}
\put (45,291) {$4s_{1/2}$}
\put (45,297) {$2g_{7/2}$}
\put (45,303) {$3d_{3/2}$}
\normalsize
\end{picture}
\vspace{15pt}
\caption[Shell model orbits]
{Single nucleon shell-model orbits and magic numbers in nuclei}
\label{orbits}
\end{figure}

The nuclear shell model has been very successful in describing 
and correlating a vast amount of experimental data. In this model 
it is assumed that each nucleon (proton or neutron) moves independently 
in a static spherical potential that represents the average interaction 
with other nucleons in the nucleus. The ordering of single nucleon levels 
(or orbits) is shown schematically in Figure~\ref{orbits}. The single  
nucleon orbits which, due to the Pauli exclusion principle, can only be  
occupied by a restricted number of identical nucleons are clustered 
into major shells. The number of protons or neutrons 
in a completely filled major shell is called a magic number. Doubly-magic 
nuclei with completely filled proton and neutron major shells are 
particularly stable. The shell model correctly reproduces all observed 
magic numbers. 

\begin{figure}
\centering
\setlength{\unitlength}{1.0pt}
\begin{picture}(100,320)(0,0)
\thicklines
\put ( 0,  0.0) {\line(1,0){60}}
\put ( 0, 77.9) {\line(1,0){60}}
\put ( 0,142.3) {\line(1,0){60}}
\put ( 0,156.7) {\line(1,0){60}}
\put ( 0,203.2) {\line(1,0){60}}
\put ( 0,215.2) {\line(1,0){60}}
\put ( 0,230) {\line(1,0){60}}
\put ( 0,242) {\line(1,0){60}}
\put ( 0,249.1) {\line(1,0){60}}
\put ( 0,253.7) {\line(1,0){60}}
\thinlines
\put (65, -5) {$2g_{9/2}$}
\put (65, 72.9) {$1i_{11/2}$}
\put (65,137.3) {$1j_{15/2}$}
\put (65,151.7) {$3d_{5/2}$}
\put (65,198.2) {$4s_{1/2}$}
\put (65,210.2) {$(3p_{1/2})^{-1}$}
\put (65,225) {}
\put (65,237) {}
\put (65,244.1) {$2g_{7/2}$}
\put (65,253.7) {$3d_{3/2}$}
\put (25,  2) {$9/2^+$}
\put (25, 79.9) {$11/2^+$}
\put (25,144.3) {$15/2^-$}
\put (25,158.7) {$5/2^+$}
\put (25,205.2) {$1/2^+$}
\put (25,217.2) {$1/2^-$}
\put (25,255.7) {$3/2^+$}
\end{picture}
\vspace{15pt}
\caption[The nucleus $^{209}\mbox{Pb}$]
{Single nucleon levels of $^{209}_{82}\mbox{Pb}_{127}$.}
\label{Pb209}
\end{figure}
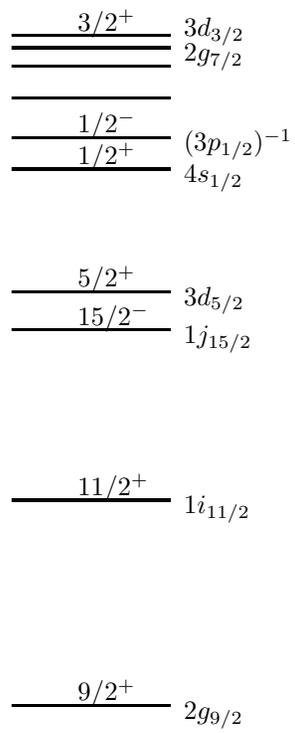

The lowest excited states of a nucleus with one nucleon outside a 
closed shell are obtained by the extra (or valence) nucleon occupying 
the various orbits in the next major shell. As an example we show 
in Figure~\ref{Pb209} the observed energy levels of the nucleus 
$^{209}_{82}\mbox{Pb}_{127}$ together with their shell model 
interpretation as a valence neutron occupying the single-particle orbits 
$2g_{9/2}$, $1i_{11/2}$, $1j_{15/2}$, $3d_{5/2}$, $4s_{1/2}$ 
$2g_{7/2}$ and $3d_{3/2}$ of the 126-184 major shell. 

The size of the model space increases rapidly if there 
are both protons and neutrons outside closed shells. 
As an example, we consider the nucleus $^{154}_{62}\mbox{Sm}_{92}$, 
which has 12 valence protons occupying the single-particle 
orbits $1g_{7/2}$, $2d_{5/2}$, $2d_{3/2}$, $3s_{1/2}$ and $1h_{11/2}$ 
of the 50-82 major shell and 10 valence neutrons occupying the 
orbits $1h_{9/2}$, $2f_{7/2}$, $2f_{5/2}$, $3p_{3/2}$, $3p_{1/2}$ 
and $1i_{13/2}$ of the 82-126 major shell. Even with the assumption 
that the lowest excited states of this nucleus can be described by 
taking into account only the valence nucleons, the shell model space 
is enormous \cite{Talmi}, as can be seen from the first column of 
Table~\ref{dim}. Despite the enormous size of the model space, the 
low-lying spectrum of $^{154}\mbox{Sm}$ shows a remarkably  
regular pattern. This suggests the existence of `effective' degrees 
of freedom, which would truncate the large shell model space 
to a manageable size, but without losing the simple features 
of the energy spectrum. 

Such an alternative is provided by the interacting boson model of 
nuclei. Its microscopic basis is the observation that the interaction 
between identical nucleons favors the formation of monopole and 
quadrupole pairs of nucleons. The interacting boson model 
can be viewed as a truncated shell model, in which 
the large shell model space spanned by the valence nucleons 
is truncated to the subspace spanned by monopole and quadrupole pairs 
of identical nucleons, which subsequently are treated as bosons. 

\subsection{The interacting boson model}

In the original formulation of the interacting boson model (IBM-1) 
no distinction is made between proton and neutron degrees 
of freedom. Low-lying collective states in even-even nuclei are 
described in terms of a system of $N$ interacting bosons 
with angular momentum and parity $L^P=0^+$ (monopole) 
and $L^P=2^+$ (quadrupole). Since the 
five components of the quadrupole boson and the monopole boson 
span a six-dimensional space with group structure $U(6)$, 
all states belong to the symmetric irreducible representation 
$[N]$ of $U(6)$, where $N$ is the total number of bosons.
In the IBM the Hamiltonian 
is expressed in second quantization. Hereto we introduce 
creation operators, $s^{\dagger}$ and $d^{\dagger}_m$, 
and annihilation operators, $s$ and $d_m$, for the bosons, 
which altogether can be denoted by $b^{\dagger}_{lm}$ and $b_{lm}$ 
with $l=0,2$ and $m=-l,-l+1,\ldots,l$ 
\ba
b^{\dagger}_{00} \;\equiv\; s^{\dagger} ~, \hspace{1cm} 
b^{\dagger}_{2m} \;\equiv\; d^{\dagger}_m ~. 
\ea
The operators $b^{\dagger}_{lm}$ and $b_{lm}$ satisfy 
standard boson commutation relations 
\ba
\, [b_{l_1m_1},b^{\dagger}_{l_2m_2}] \;=\; \delta_{l_1l_2} \delta_{m_1m_2} ~, 
\hspace{1cm} \, [b^{\dagger}_{l_1m_1},b^{\dagger}_{l_2m_2}] \;=\; 
\, [b_{l_1m_1},b_{l_2m_2}] \;=\; 0 ~. 
\ea
In second quantized form, the most general one- and two-body rotational 
invariant Hamiltonian that conserves the number of bosons is given by 
\ba
H &=& H_0 + \sum_{l} \epsilon_l \sum_m \, b^{\dagger}_{lm} b_{lm} 
+ \sum_L \sum_{l_1l_2l_3l_4} v^{(L)}_{l_1l_2l_3l_4} \, 
(b^{\dagger}_{l_1} \times b^{\dagger}_{l_2})^{(L)} \cdot 
(\tilde{b}_{l_3} \times \tilde{b}_{l_4})^{(L)} ~, \label{hibm}
\ea
with $\tilde{b}_{lm}=(-1)^{l-m} b_{l,-m}$.
The dots indicate scalar products and the 
crosses tensor products. 
Since the monopole and quadrupole bosons are identified with 
correlated pairs of valence nucleons, the number of bosons $N$ 
is determined by the total number of active proton and neutron 
pairs, counted from the nearest closed shell. 

As an example, the nucleus $^{154}_{62}$Sm$_{92}$, 
where the 12 valence protons occupy the 50-82 proton shell and 
the 10 valence neutrons occupy the 82-126 neutron shell, is treated 
in the IBM as a system of $N=6+5=11$ interacting bosons. 
The number of states with angular momentum and parity $L^P=0^+$, $2^+$ 
and $4^+$ is reduced from the shell model values by a factor 
$10^{12} - 10^{13}$ (see Table~\ref{dim}). This reduction makes 
it possible to study low-lying 
collective excitations in nuclei by diagonalizing 
Hamiltonian matrices of relatively small dimensions.  
In the last column of Table~\ref{dim} we show the dimensions of the  
model space in the neutron-proton interacting boson model (IBM-2)  
in which the neutron-proton degrees of freedom are taken into account 
explicitly. 

\subsection{Dynamical symmetries} 

In general, the Hamiltonian matrix can be diagonalized numerically 
to obtain the energy eigenvalues, but there exist also limiting 
situations in which the energy spectra can be obtained in closed 
analytic form, that is to say, in terms of an energy formula. 
These special cases correspond to dynamical symmetries, and arise 
whenever the Hamiltonian can be written in terms of Casimir invariants 
of a chain of subgroups of $U(6)$ only \cite{IBM}. 
Since nuclear states have good angular momentum, the rotation group 
$SO(3)$ in three dimensions should be included in all subgroup chains. 
Under this restriction there are three possible chains \cite{IBM} 
\ba
U(6) \supset \left\{ \begin{array}{l} 
U(5) \supset SO(5) \supset SO(3) ~, \\ \mbox{} \\
SU(3) \supset SO(3) ~, \\ \mbox{} \\
SO(6) \supset SO(5) \supset SO(3) ~. 
\end{array} \right.
\ea
The corresponding dynamical symmetries are usually referred to as the
$U(5)$, the $SU(3)$ and the $SO(6)$ limits, respectively.

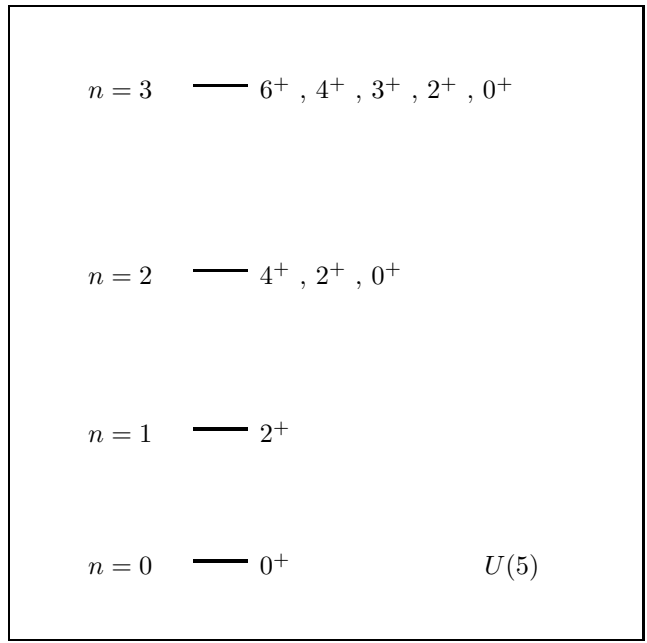
\begin{figure}
\centering
\setlength{\unitlength}{1.0pt}
\begin{picture}(240,240)(0,0)
\thinlines
\put (  0,  0) {\line(1,0){240}}
\put (  0,240) {\line(1,0){240}}
\put (  0,  0) {\line(0,1){240}}
\put (240,  0) {\line(0,1){240}}
\thicklines
\put ( 70, 30) {\line(1,0){20}}
\put ( 70, 80) {\line(1,0){20}}
\put ( 70,140) {\line(1,0){20}}
\put ( 70,210) {\line(1,0){20}}
\thinlines
\put ( 30, 25) {$n=0$}
\put ( 30, 75) {$n=1$}
\put ( 30,135) {$n=2$}
\put ( 30,205) {$n=3$}
\put ( 95, 25) {$0^+$}
\put ( 95, 75) {$2^+$}
\put ( 95,135) {$4^+ ~, \, 2^+ ~, \, 0^+$}
\put ( 95,205) {$6^+ ~, \, 4^+ ~, \, 3^+ ~, \, 2^+ ~, \, 0^+$}
\put (180, 25) {$U(5)$}
\end{picture}
\vspace{15pt}
\caption[Vibrational nucleus]
{Schematic spectrum with $U(5)$ symmetry. 
The energy levels are calculated using Eq.~(\ref{ener1}) 
with $\epsilon>0$, $\alpha>0$ and $\beta=\gamma=0$. 
The number of bosons is $N=3$.}
\label{u5}
\end{figure}

{\it (i)} In the $U(5)$ limit, the energy eigenvalues are given by 
\ba
E(n,v,L) &=& E_0 + \epsilon \, n + \alpha \, n(n+4) + \beta \, v(v+3) 
+ \gamma \, L(L+1) ~, \label{ener1}
\ea
where $n$, $v$ and $L$ are quantum numbers that label the basis states. 
Here $n$ represents the number of quadrupole bosons, $v$ is the 
boson seniority, {\it i.e.} the number of quadrupole bosons not coupled 
pairwise to angular momentum zero, and $L$ denotes the angular momentum. 
The energy spectrum is characterized by a series of multiplets labeled 
by $n$ at almost constant energy spacing 
($\alpha, \beta, \gamma \ll \epsilon$), 
which is typical for a vibrational nucleus. 
The ground state has $n=v=L=0$ and energy $E_0$. 
In Fig.~\ref{u5} we show the structure of a spectrum 
in the $U(5)$ limit. 

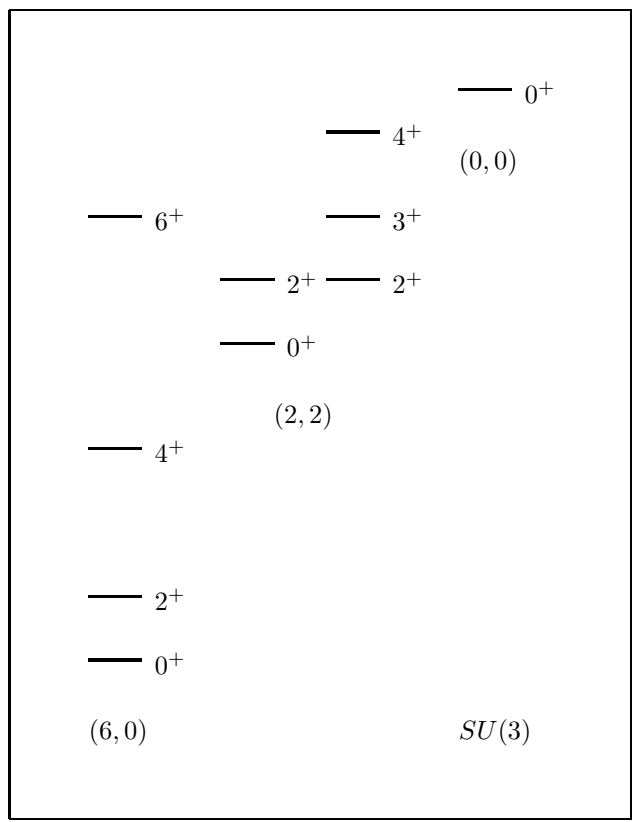
\begin{figure}
\centering
\setlength{\unitlength}{1.0pt}
\begin{picture}(235,306)(0,0)
\thinlines
\put (  0,  0) {\line(1,0){235}}
\put (  0,306) {\line(1,0){235}}
\put (  0,  0) {\line(0,1){306}}
\put (235,  0) {\line(0,1){306}}
\thicklines
\put ( 30, 60) {\line(1,0){20}}
\put ( 30, 84) {\line(1,0){20}}
\put ( 30,140) {\line(1,0){20}}
\put ( 30,228) {\line(1,0){20}}
\put ( 80,180) {\line(1,0){20}}
\put ( 80,204) {\line(1,0){20}}
\put (120,204) {\line(1,0){20}}
\put (120,228) {\line(1,0){20}}
\put (120,260) {\line(1,0){20}}
\put (170,276) {\line(1,0){20}}
\thinlines
\put ( 30, 30) {$(6,0)$}
\put ( 55, 55) {$0^+$}
\put ( 55, 79) {$2^+$}
\put ( 55,135) {$4^+$}
\put ( 55,223) {$6^+$}
\put (100,150) {$(2,2)$}
\put (105,175) {$0^+$}
\put (105,199) {$2^+$}
\put (145,199) {$2^+$}
\put (145,223) {$3^+$}
\put (145,255) {$4^+$}
\put (170,246) {$(0,0)$}
\put (195,271) {$0^+$}
\put (170, 30) {$SU(3)$}
\end{picture}
\vspace{15pt}
\caption[Rotational nucleus]
{Schematic spectrum with $SU(3)$ symmetry. 
The energy levels are calculated using Eq.~(\ref{ener2}) 
with $\kappa>0$ and $\kappa'>0$. The number of bosons is $N=3$.  
The numbers in parenthesis denote the values of $(\lambda,\mu)$.}
\label{su3}
\end{figure}

{\it (ii)} The energy eigenvalues in the $SU(3)$ limit are given by 
\ba
E(\lambda,\mu,L) &=& E_0 - \kappa \, [\lambda(\lambda+3) 
+ \mu(\mu+3) + \lambda \mu - 2N(2N+3)] + \kappa' \, L(L+1) ~. 
\label{ener2} 
\ea
Here $\lambda$, $\mu$ and $L$ label the basis states. 
The spectrum is characterized by a series of bands 
labeled by $(\lambda,\mu)$, in which the energy spacing is proportional 
to $L(L+1)$, as in a rigid rotor model. The ground state band has 
$(\lambda,\mu)=(2N,0)$ for a prolate rotor or 
$(\lambda,\mu)=(0,2N)$ for an oblate rotor. In both cases the 
ground state energy is $E_0$. 
In Fig.~\ref{su3} we show a typical spectrum in the $SU(3)$ limit. 

\begin{figure}
\centering
\setlength{\unitlength}{1.0pt}
\begin{picture}(300,290)(0,0)
\thinlines
\put (  0,  0) {\line(1,0){300}}
\put (  0,290) {\line(1,0){300}}
\put (  0,  0) {\line(0,1){290}}
\put (300,  0) {\line(0,1){290}}
\thicklines
\put ( 70, 60) {\line(1,0){20}}
\put ( 70,100) {\line(1,0){20}}
\put ( 70,160) {\line(1,0){20}}
\put ( 70,240) {\line(1,0){20}}
\put (235,220) {\line(1,0){20}}
\put (235,260) {\line(1,0){20}}
\thinlines
\put ( 70, 30) {$\sigma=3$}
\put (235,190) {$\sigma=1$}
\put ( 30, 55) {$\tau=0$}
\put ( 30, 95) {$\tau=1$}
\put ( 30,155) {$\tau=2$}
\put ( 30,235) {$\tau=3$}
\put (195,215) {$\tau=0$}
\put (195,255) {$\tau=1$}
\put ( 95, 55) {$0^+$}
\put ( 95, 95) {$2^+$}
\put ( 95,155) {$4^+ ~, \, 2^+$}
\put ( 95,235) {$6^+ ~, \, 4^+ ~, \, 3^+ ~, \, 0^+$}
\put (260,215) {$0^+$}
\put (260,255) {$2^+$}
\put (235, 30) {$SO(6)$}
\end{picture}
\vspace{15pt}
\caption[$\gamma$ unstable nucleus]
{Schematic spectrum with $SO(6)$ symmetry. 
The energy levels are calculated using Eq.~(\ref{ener3}) 
with $A>0$, $B>0$ and $C=0$. The number of bosons is $N=3$.}
\label{so6}
\end{figure}
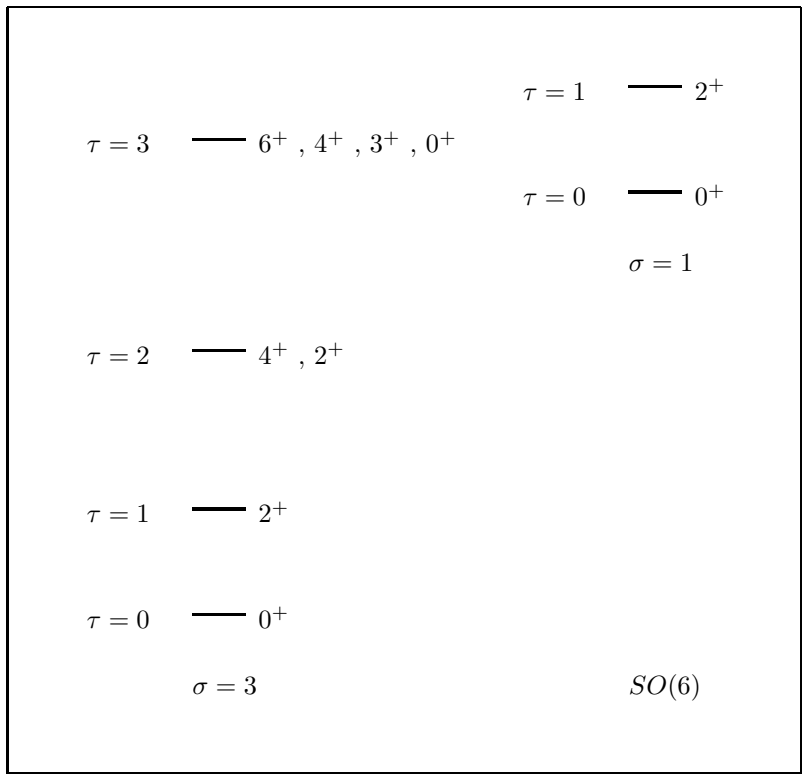

{\it (iii)} Finally, the energy formula in the $SO(6)$ limit is given by 
\ba
E(\sigma,\tau,L) &=& E_0 + A \, (N-\sigma)(N+\sigma+4) 
+ B \, \tau(\tau+3) + C \, L(L+1) ~,  
\label{ener3} 
\ea
where $\sigma$, $\tau$ and $L$ characterize the basis states. 
Here $\sigma$ and $\tau$ denote boson seniority 
labels: $\tau$ has the same meaning as $v$ in the $U(5)$ limit, 
{\it i.e.} the number of quadrupole bosons not coupled 
pairwise to angular momentum zero, whereas $\sigma$ is a generalized 
seniority that involves both monopole and quadrupole bosons. 
The energy spectrum consists of a series of vibrational multiplets 
labeled by $\sigma$, in which the energy spacing is proportional to 
the last two terms in Eq.~(\ref{ener3}).
The ground state has $\sigma=N$, $\tau=L=0$ and energy $E_0$. 
In Fig.~\ref{so6} we show a typical spectrum in the $SO(6)$ limit. 

The three dynamical symmetries provide a set of closed analytic 
expressions for energies, electromagnetic transition rates and 
selection rules that can be tested easily by experiment, and 
as such they play an important role in the qualitative 
interpretation of the data. However, only a few nuclei can be described 
by these limiting situations. We mention the low-lying states of 
$^{110}_{48}$Cd$_{62}$, $^{156}_{64}$Gd$_{92}$ and 
$^{196}_{78}\mbox{Pt}_{118}$ as good examples of nuclei with 
$U(5)$, $SU(3)$ and $SO(6)$ symmetry, respectively \cite{IBM}. 
Most nuclei display properties intermediate between the dynamical 
symmetries. In order to describe transitional regions between 
any of the three dynamical symmetries, the more general form 
of the IBM Hamiltonian of Eq.~(\ref{hibm}) has to be used. 
Its eigenvalues and eigenvectors can be obtained by 
numerical diagonalization. As examples of transitional regions 
we mention the mass region between the Pt isotopes 
and the well-deformed 
region of the rare earth nuclei, which has been interpreted in 
terms of a $SO(6) \leftrightarrow SU(3)$ transition, the Sm 
isotopes which show a sharp transition between vibrational 
and rotational spectra ($U(5) \leftrightarrow SU(3)$) and 
the Ru isotopes which show a transition between vibrational 
and $\gamma$ unstable nuclei ($U(5) \leftrightarrow SO(6)$).  

\subsection{Classical limit}

In a geometric model of collective quadrupole excitations of 
the nucleus, the nuclear surface is described by its radius 
\ba
R &=& R_0 \left[ 1 + \sum_{\mu} \alpha_{\mu} 
Y_{2\mu}(\theta,\phi) \right] ~, 
\ea
which is parametrized by five shape variables $\alpha_{\mu}$ 
($\mu=-2,\ldots,2$). Instead of $\alpha_{\mu}$ it is more convenient to 
make a transformation to the body-fixed system and to introduce the 
Hill-Wheeler coordinates $\beta$, $\gamma$ which determine the shape, 
together with the three Euler angles which determine the 
orientation in space \cite{BM}. 

The connection between the IBM and the geometric model 
can be obtained by studying the classical limit of the IBM 
by means of mean-field techniques \cite{cl}. For a system of bosons 
the variational wave function has the form of a coherent state, 
which is a condensate of $N$ deformed bosons. For static 
rotationally invariant problems, the coherent state is 
characterized by two geometric or classical variables, 
which one can associate with $\beta$ and $\gamma$. The coherent 
state is then given by  
\ba
| N;\beta,\gamma \rangle &=& \frac{1}{\sqrt{N!}} 
\left[ b_c^{\dagger}(\beta,\gamma) \right]^N \, | 0 \rangle ~, 
\ea
with 
\ba
b_c^{\dagger}(\beta,\gamma) 
&=& \frac{1}{\sqrt{1+\beta^2}} \left[ s^{\dagger} 
+ \beta \cos \gamma \, d_0^{\dagger} 
+ \frac{1}{\sqrt{2}} \beta \sin \gamma 
\left( d_2^{\dagger} + d_{-2}^{\dagger} \right) \right] ~.
\ea
For a given IBM Hamiltonian we define an energy surface 
by its expectation value in the coherent state 
\ba
E(\beta,\gamma) &\equiv& 
\langle N;\beta,\gamma \, | \, :H: \, | \, N;\beta,\gamma \rangle ~.
\ea
Taking the normal ordered product of the Hamiltonian $:H:$ amounts 
to keeping, for each interaction term, only the leading order 
contribution in the total number of bosons $N$.
For the one- and two-body Hamiltonian of Eq.~(\ref{hibm}) the 
energy surface is given by 
\ba
E(\beta,\gamma) &=& a_0 
+ \frac{N(N-1)}{(1+\beta^2)^2} \left[ a_2 \beta^2 
+ a_3 \beta^3 \cos 3\gamma + a_4 \beta^4 \right] ~, 
\ea
where the coefficients $a_i$ depend on the number of bosons $N$ and 
the parameters in the Hamiltonian. 

The classical limits of the three dynamical symmetries have 
a simple geometric interpretation. 
For the $U(5)$ limit the energy surface is given by  
\ba  
E(\beta,\gamma) &=& E_0 + \epsilon \, \frac{N\beta^2}{1+\beta^2} 
+ \alpha \, \frac{N(N-1)\beta^4}{(1+\beta^2)^2} ~, 
\ea 
whereas for the $SU(3)$ limit we find 
\ba
E(\beta,\gamma) &=& E_0 
+ \kappa \, \frac{N(N-1)}{(1+\beta^2)^2} \left[ 3\beta^4 
\mp 4\sqrt{2} \beta^3 \cos 3\gamma + 4 \right] ~, 
\ea
and for the $SO(6)$ limit 
\ba
E(\beta,\gamma) &=& E_0 
+ A \, \frac{N(N-1)}{(1+\beta^2)^2} (1-\beta^2)^2 ~.
\ea
The energy surfaces for the $U(5)$ and $SO(6)$ limits do not depend 
on the asymmetry parameter $\gamma$. For physical values of the parameters 
($\epsilon>0$ and $A >0$) they have a minimum at $\beta=0$ 
(spherical shape) and $\beta^2=1$ (deformed shape with $\gamma$ instability), 
respectively. For the $SU(3)$ limit the energy surface depends on 
both $\beta$ and $\gamma$, and has for $\kappa>0$ a minimum at 
$\beta=\sqrt{2}$ and $\gamma=0$ (axially deformed prolate shape) or at 
$\beta=\sqrt{2}$ and $\gamma=\pi/3$ (axially deformed oblate shape), 
depending on the sign of the $\beta^3 \cos 3\gamma$ term.
This analysis shows that the $U(5)$ limit corresponds to an anharmonic 
vibrator, the $SU(3)$ limit to an axial rotor with prolate or oblate 
deformation, and the $SO(6)$ limit to a $\gamma$ unstable rotor 
(or deformed oscillator). 

These results are summarized in the phase triangle of 
Fig.~\ref{shape1}, in which the equilibrium shapes corresponding 
to each one of the dynamical symmetries are 
located at the corners, and the transitional regions between 
any two of them along the three sides. Most nuclei correspond to either 
the edges or the interior of the triangle, since they are intermediate 
between two or three limiting situations. 

\section{Nucleon structure}

The nucleon itself is not an elementary particle, but a composite 
object. Effective models of the nucleon 
and its excited states (or baryon resonances) based on three 
constituents share a common spin-flavor-color structure 
but differ in their assumptions on the spatial dynamics. 
Stimulated by the success of algebraic methods in nuclear \cite{IBM} and 
molecular \cite{vibron} spectroscopy, we discuss here an interacting 
boson model for the spatial degrees of freedom \cite{BIL}. 
This model unifies various exactly solvable models of 
baryon structure, and hence provides a general framework to study the 
properties of baryon resonances in a transparent and systematic way. 

\subsection{Algebraic model of the nucleon}

Baryons are considered to be built of three constituent parts. The 
internal degrees of freedom of these three parts are taken to be:
flavor-triplet $u,d,s$ (we do not consider here heavy quark flavors), 
spin-doublet $S=1/2$, and color-triplet. The internal algebraic 
structure of the constituent parts is the usual
\ba 
{\cal G}_i &=& SU_{sf}(6) \otimes SU_{c}(3) 
\supset SU_{f}(3) \otimes SU_{s}(2) \otimes SU_{c}(3) ~. 
\label{sfc}
\ea 
In Table~\ref{baryons} we present the classification of the baryon 
flavor octet and decuplet in terms of the isospin $I$ and the 
hypercharge $Y$ according to the decomposition 
$SU_f(3) \supset SU_I(2) \otimes U_Y(1)$. The hypercharge is related 
to the electric charge $Q$ and the third component of the 
isospin $I_3$ through the Gell-Mann and Nishijima relation 
\ba
Q &=& I_3 + \frac{Y}{2} ~.
\ea
The strangeness $S$ is the difference between the hypercharge and 
the baryon number $B$ 
\ba 
S &=& Y - B ~.
\ea
The nucleon and $\Delta$ are nonstrange $S=0$, whereas the 
$\Sigma$, $\Lambda$, $\Xi$ and $\Omega$ hyperons carry strangeness 
$S=-1$, $-1$, $-2$ and $-3$, respectively. 

The relative motion of the three constituent parts is described in
terms of Jacobi coordinates, $\vec{\rho}$ and $\vec{\lambda}$,
which in the case of three identical objects are 
\ba
\vec{\rho} &=& \frac{1}{\sqrt{2}} (\vec{r}_1 - \vec{r}_2) ~,
\nonumber\\
\vec{\lambda} &=& \frac{1}{\sqrt{6}} (\vec{r}_1 + \vec{r}_2 -2\vec{r}_3) ~. 
\label{jacobi}
\ea
Here $\vec{r}_1$, $\vec{r}_2$ and $\vec{r}_3$ are the coordinates of 
the three constituents. Instead of a formulation in terms of coordinates 
and momenta we use the method of bosonic quantization, 
in which we introduce a dipole boson with $L^P=1^-$ for each independent
relative coordinate, and an auxiliary scalar boson with $L^P=0^+$ 
\cite{BIL} 
\ba 
p^{\dagger}_{\rho,m} ~, \; p^{\dagger}_{\lambda,m} ~, \; 
s^{\dagger} \hspace{1cm} (m=-1,0,1) ~. \label{bb}
\ea
The scalar boson does not represent an independent degree of freedom, 
but is added under the restriction that the total number of bosons 
$N=n_{\rho}+n_{\lambda}+n_s$ is conserved. This procedure leads to a 
compact spectrum generating algebra for the radial (or orbital) 
excitations 
\ba
{\cal G}_r &=& U(7) ~. \label{rad}
\ea
For a system of interacting bosons the model space is spanned by the 
symmetric irreducible representation $[N]$ of $U(7)$. 
The value of $N$ determines the size of the model space. 

The mass operator depends both on the spatial and the internal degrees 
of freedom. We first discuss the contribution from the spatial part, 
which is obtained by expanding the mass-squared operator $M^2$ in terms 
of the generators of $U(7)$ \cite{BIL} similar to Eq.~(\ref{hibm}), 
but now the boson operators $b^{\dagger}_{lm}$ and $b_{lm}$ can be 
any one of building blocks of Eq.~(\ref{bb}). Because of parity 
conservation only interaction terms with an even number of dipole 
boson operators are permitted. 
For nonstrange $qqq$ baryons, the mass-squared operator $M^2$ 
has to be invariant under the permuation group $S_3$, {\it i.e.} 
under the interchange of any of the three constituent parts. 
This poses an additional constraint on the allowed interaction terms. 
The wave functions have, by construction, good 
angular momentum $L$, parity $P$, and permutation symmetry $t$. 
The three symmetry classes of the $S_3$ permutation group are
characterized by the irreducible representations: 
$t=S$ for the one-dimensional symmetric representation, 
$t=A$ for the one-dimensional antisymmetric representation,  
and $t=M$ for the two-dimensional mixed symmetry representation. 

\subsection{Dynamical symmetries} 

The $S_3$ invariant $U(7)$ mass operator has a rich group structure. 
Just as in the case of the interacting boson model for nuclei, 
it is of general interest to study limiting situations, in which 
the mass spectrum can be obtained in closed form. These special 
solutions correspond to dynamical symmetries of the model. 
Under the restriction that the eigenstates have good angular momentum, 
parity and permutation symmetry, there are several possibilities. 
Here we consider the chains 
\ba
U(7) \supset \left\{ \begin{array}{l} 
U(6) \supset \left\{ \begin{array}{l} {\cal SU}(3) \otimes SU(2) 
\supset {\cal SO}(3) \otimes SO(2) ~, \\ \mbox{} \\
SO(6) \supset SU(3) \otimes SO(2) 
\supset {\cal SO}(3) \otimes SO(2) ~, \end{array} \right. \\ \mbox{} \\
SO(7) \supset SO(6) \supset SU(3) \otimes SO(2) 
\supset {\cal SO}(3) \otimes SO(2) ~.
\end{array} \right. 
\label{lattice} 
\ea
The corresponding dynamical symmetries are referred to as the
$U(6) \supset {\cal SU}(3) \otimes SU(2)$ limit, the $U(6) \supset SO(6)$ 
limit and the $SO(7)$ limit, respectively. 
These chains have the direct product group ${\cal SO}(3) \otimes SO(2)$ 
in common, where ${\cal SO}(3)$ is the angular momentum group and 
$SO(2)$ is related to the permutation symmetry \cite{BIL,BDL,KM}. 

{\it (i)} The first chain corresponds to the problem of three particles 
in a common harmonic oscillator potential \cite{KM}. It separates 
the behavior in three-dimensional coordinate space determined by 
${\cal SU}(3) \supset {\cal SO}(3)$, from that in the index space, 
given by $SU(2) \supset SO(2)$. In this limit 
the eigenvalues are given by 
\ba
M^2(n,L,F,M_F) &=& M^2_0 + \epsilon_1 \, n + \epsilon_2 \, n(n+5) 
\nonumber\\
&&+ \alpha \, F(F+2) + \kappa \, L(L+1) + \kappa^{\prime} M_F^2 ~. 
\label{mass1}
\ea
Fig.~\ref{anhosc} shows the structure of a spectrum 
with $U(6)$ symmetry. 
The levels are grouped into oscillator shells characterized by $n$. 
The ground state has $n=0$ and $L^P_t=0^+_{S}$. The one-phonon 
multiplet $n=1$ has two degenerate states with $L^P=1^-$ which 
belong to the two-dimensional representation $M$ of the permutation group, 
and the two-phonon multiplet $n=2$ consists of the states 
$L^P_t=2^+_{S}$, $2^+_{M}$, $1^+_{A}$, $0^+_{S}$ and $0^+_{M}$. 
The splitting within an oscillator shell is determined by the last 
three terms of Eq.~(\ref{mass1}). 

{\it (ii)} Another classification scheme for the six-dimensional 
oscillator is provided by the second group chain of Eq.~(\ref{lattice}). 
The reduction $U(6) \supset SO(6) \supset SU(3) \otimes SO(2)$ 
has been studied in detail in \cite{Chacon}. Here it is embedded 
in $U(7)$. The spectrum of the $U(6) \supset SO(6)$ limit 
is given by 
\ba
M^2(n,\sigma,L,M_F) &=& M^2_0 + \epsilon_1 \, n + \epsilon_2 \, n(n+5) 
\nonumber\\ 
&&+ \beta \, \sigma(\sigma+4) + \kappa \, L(L+1) + \kappa^{\prime} M_F^2 ~. 
\label{mass2}
\ea
Also in this case the levels are grouped into oscillator shells 
according to Fig.~\ref{anhosc}. However, in this case the splitting 
within an oscillator shell which is determined by the last 
three terms of Eq.~(\ref{mass2}) is different from 
that in the $U(6) \supset {\cal SU}(3) \otimes SU(2)$ limit. 

\begin{figure}
\centering
\setlength{\unitlength}{1.0pt}
\begin{picture}(300,270)(0,0)
\thinlines
\put (  0,  0) {\line(1,0){300}}
\put (  0,270) {\line(1,0){300}}
\put (  0,  0) {\line(0,1){270}}
\put (300,  0) {\line(0,1){270}}
\thicklines
\put ( 70, 60) {\line(1,0){20}}
\put ( 70,110) {\line(1,0){20}}
\put ( 70,230) {\line(1,0){20}}
\put (235,200) {\line(1,0){20}}
\thinlines
\put ( 70, 30) {$\omega=2$}
\put (235,170) {$\omega=0$}
\put ( 30, 55) {$\sigma=0$}
\put ( 30,105) {$\sigma=1$}
\put ( 30,225) {$\sigma=2$}
\put (195,195) {$\sigma=0$}
\put ( 95, 55) {$0^+_S$}
\put ( 95,105) {$1^-_M$}
\put ( 95,225) {$2^+_S ~, \, 2^+_M ~, \, 1^+_A ~, \, 0^+_M$}
\put (260,195) {$0^+_S$}
\put (235, 30) {$SO(7)$}
\end{picture}
\vspace{15pt}
\caption[Deformed oscillator]
{Schematic spectrum with $SO(7)$ symmetry. 
The masses are calculated using Eq.~(\ref{mass3}) 
with $A>0$, $\beta>0$ and $\kappa=\kappa^{\prime}=0$. 
The number of bosons is $N=2$.}
\label{defosc}
\end{figure}
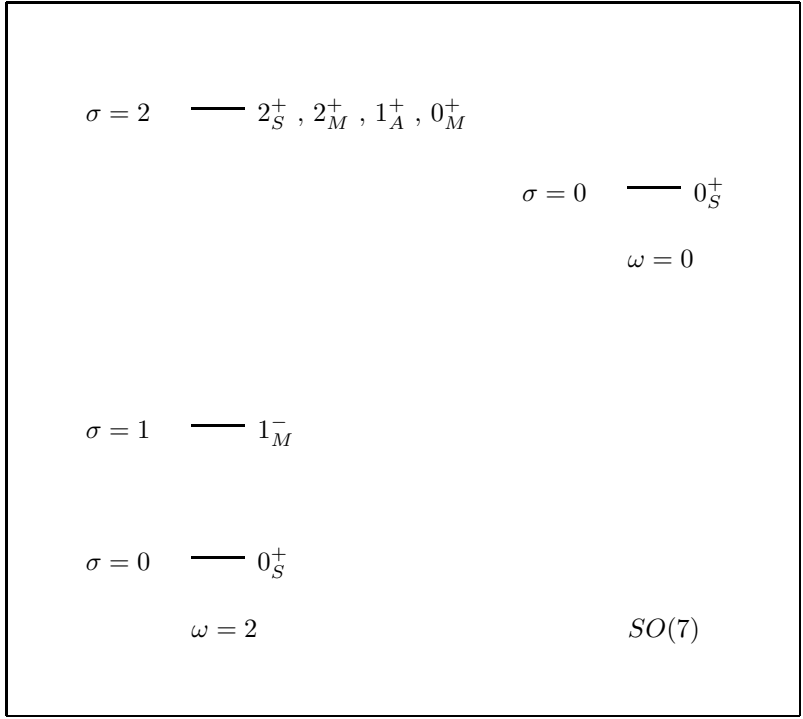

{\it (iii)} The two group chains associated with the $U(7) \supset U(6)$ 
reduction correspond a six-dimensional anharmonic oscillator, for 
which the total number of oscillator quanta $n$ is a good 
quantum number. However, this is no longer the case for the third 
dynamical symmetry of Eq.~(\ref{lattice}). 
In the $SO(7)$ limit the eigenvalues are 
\ba
M^2(\omega,\sigma,L,M_F) &=& M^2_0 + A \, (N-\omega)(N+\omega+5) 
\nonumber\\ 
&& + \beta \, \sigma(\sigma+4) + \kappa \, L(L+1) + \kappa^{\prime} M_F^2 ~. 
\label{mass3}
\ea
Analogous to the $SO(6)$ limit of the IBM, the $SO(7)$  
limit corresponds to a deformed oscillator. 
In Fig.~\ref{defosc} we show a typical spectrum with $SO(7)$ symmetry. 
The states are now ordered in bands characterized by 
$\omega$, rather than in harmonic oscillator shells, as in the 
previous two examples. 

\subsection{Classical limit}

A more intuitive geometric interpretation of algebraic $U(7)$ 
interactions can be obtained by studying its classical limit. 
The procedure is similar to that discussed in Section 2.3 for the 
interacting boson model of nuclei. 
The coherent state is a condensate of $N$ deformed bosons, which 
for static rotationally invariant problems can be parametrized as 
\ba
b_c^{\dagger}(r,\chi,\theta) &=& \frac{1}{\sqrt{1+r^2}} 
\left[ s^{\dagger} + r \cos \chi \, p_{\lambda,x}^{\dagger}
+ r \sin \chi \, (\cos \theta \, p_{\rho,x}^{\dagger} 
+ \sin \theta \, p_{\rho,y}^{\dagger}) \right] ~. 
\label{bc} 
\ea
The geometry is chosen such that $\vec{\rho}$ and $\vec{\lambda}$ 
span the $xy$ plane with the $x$-axis along 
$\vec{\lambda}$ and the $z$-axis perpendicular to this plane. 
The two vectors $\vec{\rho}$ and $\vec{\lambda}$ 
are parametrized in terms of the three Euler angles which 
are associated with the orientation of the system, 
and three internal coordinates which are taken as the two lengths 
of the vectors $r_{\lambda} = r \cos \chi$ and $r_{\rho} = r \sin \chi$, 
and their relative angle $\theta$. 
The hyperradius $r$ is a measure of the dimension of the system, 
whereas the hyperangle $\chi$ and the angle $\theta$ determine its 
shape \cite{hyper}. The surface associated with one- and 
two-body $S_3$ invariant interactions is given by 
\ba
M^2(r,\chi,\theta) &=& a_0 
+ \frac{N(N-1)}{(1+r^2)^2} \left[ a_2 r^2 + a_4 r^4 
- b r^4 \sin^2 (2\chi) \sin^2 \theta \right] ~.
\label{es}
\ea
The coefficients $a_i$ and $b$ depend on the number of 
bosons $N$ and the parameters in the mass-squared operator. 

The classical limits of the three dynamical symmetries have 
a simple geometric interpretation. 
For the $U(6) \supset {\cal SU}(3) \otimes SU(2)$ limit 
the surface is given by 
\ba 
M^2(r,\chi,\theta) &=& M^2_0 + \epsilon_1 \, \frac{Nr^2}{1+r^2} 
+ \epsilon_2 \, \frac{N(N-1)r^4}{(1+r^2)^2} 
\nonumber\\
&&+ \alpha \frac{N(N-1)r^4}{(1+r^2)^2} 
[ 1 - \sin^2 (2\chi) \sin^2 \theta ] ~,  
\ea 
whereas for the $U(6) \supset SO(6)$ limit we find 
\ba
M^2(r,\chi,\theta) &=& M^2_0 + \epsilon_1 \, \frac{Nr^2}{1+r^2} 
+ \epsilon_2 \, \frac{N(N-1)r^4}{(1+r^2)^2} ~, 
\ea
and for the $SO(7)$ limit  
\ba
M^2(r,\chi,\theta) &=& M^2_0 
+ A \, \frac{N(N-1)}{(1+r^2)^2} (1-r^2)^2 ~.
\ea
The surfaces for the $U(6) \supset SO(6)$ and $SO(7)$ limits do 
not depend on the angles $\chi$ and $\theta$. For physical values of the 
parameters ($\epsilon_1>0$ and $A > 0$) they have 
a minimum at $r=0$ (spherical shape) and $r^2=1$ (deformed shape with 
$\chi$ and $\theta$ instability), respectively. 
For the $U(6) \supset {\cal SU}(3) \otimes SU(2)$ limit 
the surface depends on all three geometric variables, 
the radius $r$ and the angles $\chi$ and $\theta$. For realistic 
values of the parameters ($\epsilon_1 > 0$) the minimum is at $r=0$ 
(spherical shape), just as for the $U(6) \supset SO(6)$ limit. 
This analysis shows that the two $U(6)$ limits correspond 
to an anharmonic vibrator, and the $SO(7)$ limit to a deformed 
oscillator (or $\chi$, $\theta$ unstable rotor).  

\begin{figure} 
\centering
\vspace{2cm}
\setlength{\unitlength}{1pt}
\begin{picture}(400,200)
\thinlines
\put ( 85, 50) {$v_1$}
\put ( 50,100) {\circle*{10}}
\put (130,100) {\circle*{10}}
\put ( 90,180) {\circle*{10}}
\put ( 50,100) {\line ( 4, 3){40}} 
\put (130,100) {\line (-4, 3){40}} 
\put ( 90,180) {\line ( 0,-1){50}} 
\thicklines
\put ( 50,100) {\vector(-4,-3){12}}
\put (130,100) {\vector( 4,-3){12}}
\put ( 90,180) {\vector( 0, 1){15}}
\thinlines
\put (215, 50) {$v_{2a}$}
\put (180,100) {\circle*{10}}
\put (260,100) {\circle*{10}}
\put (220,180) {\circle*{10}}
\put (180,100) {\line ( 4, 3){40}} 
\put (260,100) {\line (-4, 3){40}} 
\put (220,180) {\line ( 0,-1){50}} 
\thicklines
\put (180,100) {\vector( 4,-3){12}}
\put (260,100) {\vector(-4,-3){12}}
\put (220,180) {\vector( 0, 1){15}}
\thinlines
\put (345, 50) {$v_{2b}$}
\put (310,100) {\circle*{10}}
\put (390,100) {\circle*{10}}
\put (350,180) {\circle*{10}}
\put (310,100) {\line ( 4, 3){40}} 
\put (390,100) {\line (-4, 3){40}} 
\put (350,180) {\line ( 0,-1){50}} 
\thicklines
\put (310,100) {\vector(-1,-2){ 6}}
\put (390,100) {\vector(-1, 2){ 6}}
\put (350,180) {\vector( 1, 0){15}}
\end{picture}
\caption[Vibrations]{Fundamental vibrations of X$_3$ configuration.}
\label{vib}
\end{figure}
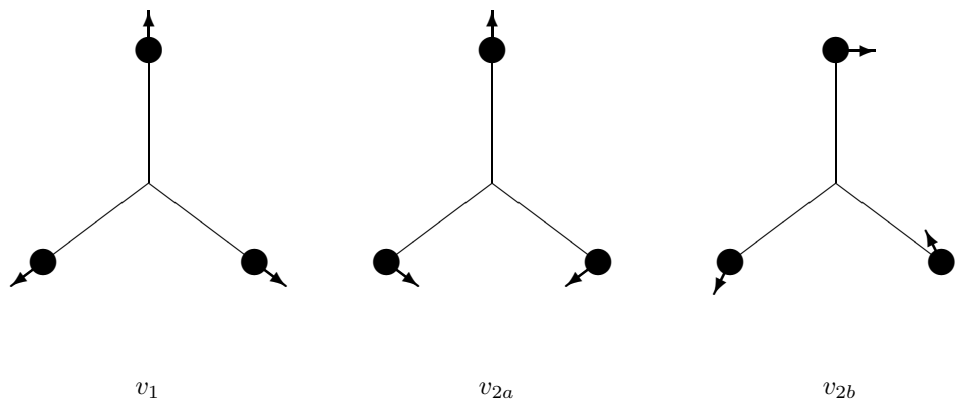

It is interesting to note that the surface of Eq.~(\ref{es}) 
has another equilibrium shape, that does not correspond 
to one of the dynamical symmetries discussed above. 
We consider the operator \cite{BIL,BDL} 
\ba
M^2 &=& \xi_1 \, ( R^2 \, s^{\dagger} s^{\dagger}
- p^{\dagger}_{\rho} \cdot p^{\dagger}_{\rho}
- p^{\dagger}_{\lambda} \cdot p^{\dagger}_{\lambda} ) \,
( R^2 \, \tilde{s} \tilde{s} - \tilde{p}_{\rho} \cdot \tilde{p}_{\rho}
- \tilde{p}_{\lambda} \cdot \tilde{p}_{\lambda} )
\nonumber\\
&& + \xi_2 \, \left[ ( p^{\dagger}_{\rho} \cdot p^{\dagger}_{\rho}
- p^{\dagger}_{\lambda} \cdot p^{\dagger}_{\lambda} ) \,
( \tilde{p}_{\rho} \cdot \tilde{p}_{\rho}
- \tilde{p}_{\lambda} \cdot \tilde{p}_{\lambda} )
+ 4 \, ( p^{\dagger}_{\rho} \cdot p^{\dagger}_{\lambda} ) \,
( \tilde{p}_{\lambda} \cdot \tilde{p}_{\rho} ) \right] ~.
\label{oblate}
\ea
For $R^2=0$, the mass-squared operator of 
Eq.~(\ref{oblate}) has $U(7) \supset U(6)$ symmetry and corresponds
to an anharmonic vibrator, whereas for $R^2=1$ and $\xi_2=0$ it has 
$U(7) \supset SO(7)$ symmetry and corresponds to a deformed 
oscillator. The general case with $R^2 \neq 0$ and $\xi_1$, $\xi_2>0$ 
corresponds to an oblate symmetric top \cite{BIL,BDL}. This can be seen 
by studying the classical limit and performing a normal mode analysis. 
The corresponding surface 
\ba 
M^2(r,\chi,\theta) &=& 
\xi_1 \, \frac{N(N-1)}{(1+r^2)^2} (R^2-r^2)^2 
+ \xi_2 \, \frac{N(N-1)r^4}{(1+r^2)^2} 
[ 1 - \sin^2 (2\chi) \sin^2 \theta ] ~, 
\ea 
has a stable nonlinear equilibrium shape characterized by 
$r=R$, $\chi=\pi/4$ and $\theta=\pi/2$, {\it i.e.} the two coordinates 
have equal length and are perpendicular. 
These two conditions are precisely those satisfied by the Jacobi
coordinates of Eq.~(\ref{jacobi}) for an equilateral triangle. 
In a normal mode analysis, the mass-squared operator of Eq.~(\ref{oblate}) 
reduces to leading order in $N$ to a harmonic form,  
and its spectrum is given by \cite{BIL,BDL} 
\ba
M^2(v_1,v_2) &=& \kappa_1 \, v_1 + \kappa_2 \, v_2 ~, 
\label{evib}
\ea
with 
\ba
\kappa_1 &=& \xi_1 \, 4 N R^2 ~,
\nonumber\\
\kappa_2 &=& \xi_2 \, 4 N R^2/(1+R^2) ~. 
\ea
Here $v_1$ represents the number of quanta in a symmetric stretching 
vibration, and $v_2=v_{2a}+v_{2b}$ denotes the total number of quanta 
in a degenerate doublet which consists of an antisymmetric 
stretching vibration ($v_{2a}$) and a bending vibration ($v_{2b}$). 
This pattern is in agreement with the point-group classification of 
the fundamental vibrations of a symmetric $X_3$ configuration 
\cite{Herzberg} (see Fig.~\ref{vib}). 
Therefore, the condensate boson of Eq.~(\ref{bc}) with 
$r=R$, $\chi=\pi/4$ and $\theta=\pi/2$, 
corresponds to the geometry of an oblate symmetric top 
with the threefold symmetry axis along the $z$-axis. 

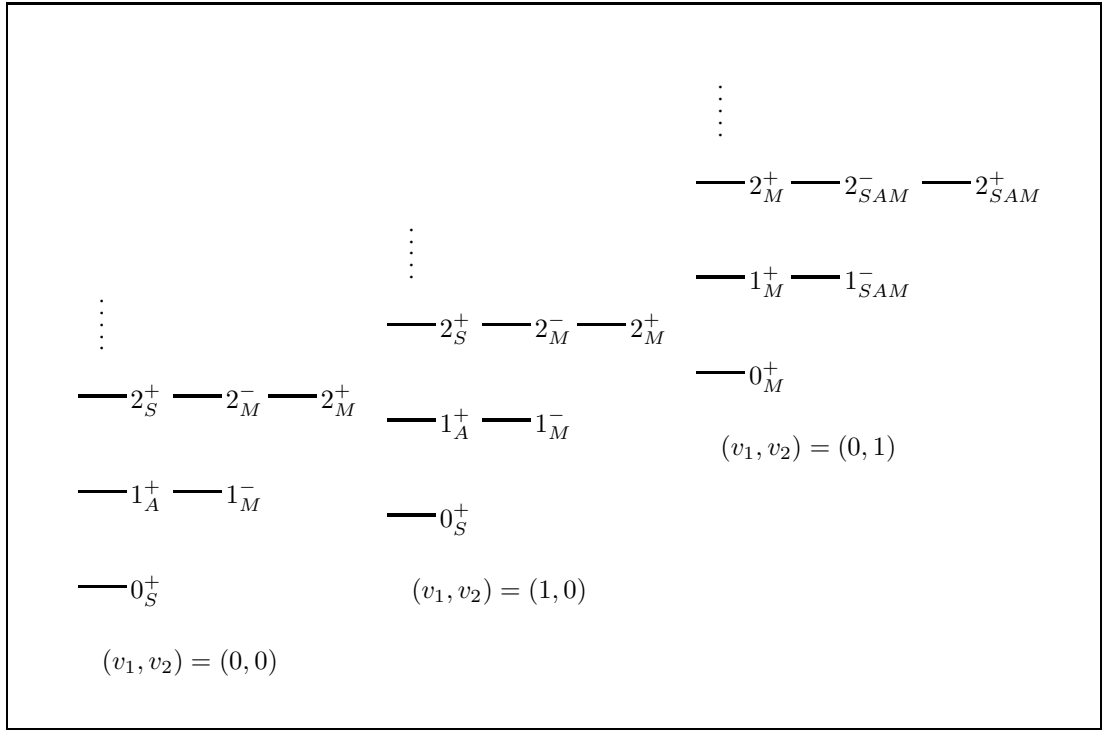
\begin{figure}
\centering
\setlength{\unitlength}{0.9pt}
\begin{picture}(460,305)(0,0)
\thinlines
\put (  0,  0) {\line(1,0){460}}
\put (  0,305) {\line(1,0){460}}
\put (  0,  0) {\line(0,1){305}}
\put (460,  0) {\line(0,1){305}}
\thicklines
\put ( 30, 60) {\line(1,0){20}}
\put ( 30,100) {\line(1,0){20}}
\put ( 30,140) {\line(1,0){20}}
\put ( 70,100) {\line(1,0){20}}
\put ( 70,140) {\line(1,0){20}}
\put (110,140) {\line(1,0){20}}
\multiput ( 40,160)(0,5){5}{\circle*{0.1}}
\thinlines
\put ( 40, 25) {$(v_1,v_2)=(0,0)$}
\put ( 52, 55) {$0^+_S$}
\put ( 52, 95) {$1^+_A$}
\put ( 92, 95) {$1^-_M$}
\put ( 52,135) {$2^+_S$}
\put ( 92,135) {$2^-_M$}
\put (132,135) {$2^+_M$}
\thicklines
\put (160, 90) {\line(1,0){20}}
\put (160,130) {\line(1,0){20}}
\put (160,170) {\line(1,0){20}}
\put (200,130) {\line(1,0){20}}
\put (200,170) {\line(1,0){20}}
\put (240,170) {\line(1,0){20}}
\multiput (170,190)(0,5){5}{\circle*{0.1}}
\thinlines
\put (170, 55) {$(v_1,v_2)=(1,0)$}
\put (182, 85) {$0^+_S$}
\put (182,125) {$1^+_A$}
\put (222,125) {$1^-_M$}
\put (182,165) {$2^+_S$}
\put (222,165) {$2^-_M$}
\put (262,165) {$2^+_M$}
\thicklines
\put (290,150) {\line(1,0){20}}
\put (290,190) {\line(1,0){20}}
\put (290,230) {\line(1,0){20}}
\put (330,190) {\line(1,0){20}}
\put (330,230) {\line(1,0){20}}
\put (385,230) {\line(1,0){20}}
\multiput (300,250)(0,5){5}{\circle*{0.1}}
\thinlines
\put (300,115) {$(v_1,v_2)=(0,1)$}
\put (312,145) {$0^+_M$}
\put (312,185) {$1^+_M$}
\put (352,185) {$1^-_{SAM}$}
\put (312,225) {$2^+_M$}
\put (352,225) {$2^-_{SAM}$}
\put (407,225) {$2^+_{SAM}$}
\end{picture}
\vspace{1cm}
\caption[Oblate top]
{Schematic spectrum of an oblate symmetric top. 
The masses are calculated using Eq.(\ref{mass4}) with 
$\kappa_1>0$, $\kappa_2>0$ and $\alpha>0$.}
\label{top}
\end{figure}

In Fig.~\ref{top} we show a schematic spectrum of an oblate 
symmetric top. In anticipation of the application to the 
mass spectrum of nonstrange baryon resonances we have added 
a term linear in the angular momentum $L$. 
The spectrum consists of a series of vibrational excitations 
characterized by the labels ($v_1,v_2$), and a tower of 
rotational excitations built on top of each vibration. 

The results of the analysis of the classical limit of $S_3$ 
invariant one- and two-body interactions in $U(7)$ 
are summarized in the phase triangle of Fig.~\ref{shape2}, 
in which the three equilibrium shapes are located at the corners. 
This phase triangle is very similar as the one for the nuclear case: 
there is a spherical shape, a deformed shape that does not depend 
on the angular variables, and one rigid deformed shape. 
An important difference is that, whereas in the nuclear case there 
exists a large amount of collective nuclei which either correspond 
to one of the dynamical symmetries or to a transitional region 
between them, in the nucleon case there is only one single 
baryon spectrum. The question is now: if we assume that the 
radial excitations of the nucleon can be described by $U(7)$, 
where does the nonstrange baryon mass spectrum fit in this triangle? 

\subsection{Nonstrange baryons}

Here we study the mass spectrum of the nonstrange baryon 
resonances of the nucleon 
(isospin $I=1/2$) and the delta (isospin $I=3/2$) family. 
The radial excitations are described in terms 
of the $U(7)$ interacting boson model which was discussed in the 
previous sections. The full algebraic structure is obtained by 
combining the radial part of Eq.~(\ref{rad}) with the internal 
spin-flavor-color part of Eq.~(\ref{sfc})
\ba
{\cal G} \;=\; {\cal G}_r \otimes {\cal G}_i 
\;=\; U(7) \otimes SU_{sf}(6) \otimes SU_c(3) ~.
\ea
The spatial part of the baryon wave function has to be combined 
with the spin-flavor and color part, in such a way that the total 
wave function is antisymmetric. Since the color part of the wave 
function is antisymmetric (color singlet), the remaining part 
(spatial plus spin-flavor) has to be symmetric. For nonstrange 
resonances which have three identical constituent parts this means 
that the symmetry of the spatial wave function under $S_3$ is 
the same as that of the spin-flavor part. Therefore, one can use the 
representations of either $S_3$ or $SU_{sf}(6)$ to label the states. 
The subsequent decomposition of representations of 
$SU_{sf}(6)$ into those of $SU_{f}(3) \otimes SU_{s}(2)$
is the standard one
\ba
S \;\leftrightarrow\; [56] &\supset& ^{2}8 \, \oplus \, ^{4}10 ~,
\nonumber\\
M \;\leftrightarrow\; [70] &\supset& 
^{2}8 \, \oplus \, ^{4}8 \, \oplus \, ^{2}10 \, \oplus \, ^{2}1 ~,
\nonumber\\
A \;\leftrightarrow\; [20] &\supset& ^{2}8 \, \oplus \, ^{4}1 ~. 
\label{usf6}
\ea
Here the representations of the spin-flavor groups $SU_{sf}(6)$, 
$SU_f(3)$ and $SU_s(2)$ are denoted by their dimensions. 
The total baryon wave function is expressed as 
\ba
\left| \Psi \right> &=& \left| \, ^{2S+1}\mbox{dim}\{SU_f(3)\}_J \, 
[\mbox{dim}\{SU_{sf}(6)\},L^P] \, \right> ~,
\label{baryonwf}
\ea
where $S$ and $J$ are the spin and total angular momentum 
$\vec{J}=\vec{L}+\vec{S}$~. The ground state baryons of Table~\ref{baryons} 
have $L^P_t=0^+_S$, and are labeled by 
$| \, ^{2}8_{1/2} \, [56,0^+] \, \rangle$ for the $J^P=1/2^+$ octet 
and $| \, ^{4}10_{3/2} \, [56,0^+] \, \rangle$ for the $J^P=3/2^+$ 
decuplet. 

We analyze the mass spectrum of nonstrange baryon resonances 
in terms of the mass formula 
\ba
M^2 &=& M^2_0 + M^2_{radial} + M^2_{sf} ~.
\ea
The radial excitations of the nucleon are interpreted as 
vibrations and rotations of an oblate symmetric top \cite{BIL} 
\ba
M^2_{radial} &=& \kappa_1 \, v_1 + \kappa_2 \, v_2 + \alpha \, L ~. 
\label{mass4}
\ea
The N(1440) and N(1710) resonances are 
associated with vibrational excitations with $(v_1,v_2)=(1,0)$ and 
$(0,1)$, respectively. 
The spin-flavor contribution to the mass-squared operator is 
expressed in a G\"ursey-Radicati form \cite{GR} 
\ba
M^2_{sf}&=& a \, \left[ \langle C_{2SU_{sf}(6)} \rangle - 45 \right] 
+ b \, \left[ \langle C_{2SU_{f}(3)} \rangle -  9 \right]
+ c \, \left[ \langle C_{2SU_{s}(2)} \rangle - \frac{3}{4} \right] ~. 
\ea
According to Eq.~(\ref{usf6}), the 
$SU_{sf}(6)$ term depends on the permutation symmetry of 
the wave functions. The $SU_{f}(3)$ term only depends on the flavor, 
and the $SU_{s}(2)$ term contains the spin dependence.
A simultaneous fit to 25 well-established (3 and 4 star) nucleon 
and delta resonances gives a r.m.s. deviation of 39 MeV \cite{BIL}. 
In Table~\ref{barres} we show all calculated resonances below 2 GeV. 
Especially in the nucleon sector there are many more states 
calculated than have been observed so far. The lowest socalled 
`missing' resonances correspond to the unnatural parity states 
with $L^P=1^+$, $2^-$, which are decoupled both in electromagnetic 
and strong decays, and hence very difficult to observe. The resonances 
in square brackets are not very well established experimentally 
(1 and 2 star) and are tentatively assigned as candidates for some 
of the missing states. 

\section{Summary and conclusions}

In this contribution we have discussed interacting boson models of 
nuclear and nucleon structure. Although the energy scales involved 
in the two applications differ by three orders of magnitude 
(several MeV's for nuclear excitations and several GeV's for 
excitations of the nucleon), in both cases such algebraic models 
provide an elegant and, at the same time, powerful method to 
describe collective excitations of complex systems by introducing 
a set of effective degrees of freedom. 

There are two advantages to these type of models that are worth 
mentioning. First of all, the use of algebraic techniques 
makes it straightforward to obtain eigenvalues and eigenvectors. 
This is done by means of matrix diagonalization, rather than by 
solving a set of coupled differential equations. Secondly, 
the existence of dynamical symmetries makes it possible to derive 
closed analytic expressions for energies, electromagnetic transition 
rates, decay widths and selection rules that can be tested easily 
by experiment, and as such they play an important role in the 
qualitative interpretation of the data. 

A geometric interpretation of algebraic interactions has been obtained 
by studying its classical limit. This way it was shown that interacting 
boson models unify various exactly solvable models in a single 
framework. In the nuclear case, we showed that the three dynamical 
symmetries correpond to the anharmonic vibrator, the axially deformed 
rotor and the $\gamma$ unstable rotor, respectively. For nonstrange 
baryon resonances, the $U(7)$ model contains 
the (an)harmonic oscillator, the deformed oscillator and 
the oblate symmetric top as special limiting cases. 

In conclusion, interacting boson models provide a general framework 
to study collective excitations of complex systems in a 
transparent and systematic way. 

\section*{Acknowledgements}

This work is supported in part by DGAPA-UNAM under project IN101997,
and by grant No. 94-00059 from the United States-Israel Binational
Science Foundation.

\clearpage

\begin{table}
\centering
\caption[Dimensions]{Model space for $^{154}_{62}$Sm$_{92}$.}
\label{dim}
\vspace{15pt}
\begin{tabular}{crcc}
\hline
& & & \\
$L^P$ & Shell model \cite{Talmi} & IBM-1 & IBM-2 \\
& & & \\
\hline
& & & \\
$0^+$ &  41,654,193,517,797 & 16 & 204 \\
$2^+$ & 346,132,052,934,889 & 26 & 680 \\
$4^+$ & 530,897,397,260,575 & 30 & 934 \\
& & & \\
\hline
\end{tabular}
\end{table}

\clearpage

\begin{table}
\centering
\caption[Baryons]{Classification of ground state baryons 
according to $SU_f(3) \supset SU_I(2) \otimes U_Y(1)$.} 
\label{baryons} 
\vspace{15pt} 
\begin{tabular}{lllcrc}
\hline
& & & & & \\
& & & $I$ & $Y$ & $Q$ \\
& & & & & \\
\hline
& & & & & \\
$J^P=\frac{1}{2}^+$ octet 
& Nucleon & $N$       & $\frac{1}{2}$ &   1 &     0,1 \\
& Sigma   & $\Sigma$  &  1  &   0 & --1,0,1 \\
& Lambda  & $\Lambda$ &  0  &   0 &     0 \\
& Xi      & $\Xi$     & $\frac{1}{2}$ & --1 & --1,0 \\
& & & & & \\
\hline
& & & & & \\
$J^P=\frac{3}{2}^+$ decuplet 
& Delta & $\Delta$        & $\frac{3}{2}$ &   1 & --1,0,1,2 \\
& Sigma & $\Sigma^{\ast}$ &  1  &   0 & --1,0,1 \\
& Xi    & $\Xi^{\ast}$    & $\frac{1}{2}$ & --1 & --1,0 \\
& Omega & $\Omega$        &  0  & --2 & --1 \\
& & & & & \\
\hline
\end{tabular}
\end{table}

\clearpage

\begin{table}
\centering
\caption[Nonstrange baryons]{
All calculated nucleon and delta resonances (in MeV) below 2 GeV. 
Tentative assignments of 1 and 2 star resonances \cite{PDG} are shown
in brackets.}
\label{barres} 
\vspace{15pt} 
\begin{tabular}{cccl}
\hline
& & & \\
State & ($v_1,v_2$) & $M_{\mbox{calc}}$ & Baryon \\
& & & \\
\hline
& & & \\
$^{2}8_J[56,0^+]$ & (0,0) &  939 & N$(939)P_{11}$ \\
$^{2}8_J[70,1^-]$ & (0,0) & 1566 & 
N$(1535)S_{11}$, N$(1520)D_{13}$ \\
$^{4}8_J[70,1^-]$ & (0,0) & 1680 & 
N$(1650)S_{11}$, N$(1700)D_{13}$, N$(1675)D_{15}$ \\
$^{2}8_J[20,1^+]$ & (0,0) & 1720 & \\
$^{2}8_J[56,2^+]$ & (0,0) & 1735 & 
N$(1720)P_{13}$, N$(1680)F_{15}$ \\
$^{2}8_J[70,2^-]$ & (0,0) & 1875 & \\
$^{2}8_J[70,2^+]$ & (0,0) & 1875 & 
[N$(1900)P_{13}$], [N$(2000)F_{15}$] \\
$^{4}8_J[70,2^-]$ & (0,0) & 1972 & \\
$^{4}8_J[70,2^+]$ & (0,0) & 1972 & [N$(1990)F_{17}$] \\
$^{2}8_J[56,0^+]$ & (1,0) & 1440 & N$(1440)P_{11}$ \\
$^{2}8_J[70,1^-]$ & (1,0) & 1909 & \\
$^{2}8_J[70,0^+]$ & (0,1) & 1710 & N$(1710)P_{11}$ \\
$^{4}8_J[70,0^+]$ & (0,1) & 1815 & \\
$^{2}8_J[56,1^-]$ & (0,1) & 1866 & \\
$^{2}8_J[70,1^+]$ & (0,1) & 1997 & \\
$^{2}8_J[70,1^-]$ & (0,1) & 1997 & \\
& & & \\
\hline
& & & \\
$^{4}10_J[56,0^+]$ & (0,0) & 1232 & $\Delta(1232)P_{33}$ \\
$^{2}10_J[70,1^-]$ & (0,0) & 1649 & 
$\Delta(1620)S_{31}$, $\Delta(1700)D_{33}$ \\
$^{4}10_J[56,2^+]$ & (0,0) & 1909 & $\Delta(1910)P_{31}$, 
$\Delta(1920)P_{33}$, $\Delta(1905)F_{35}$, $\Delta(1950)F_{37}$ \\
$^{2}10_J[70,2^-]$ & (0,0) & 1945 & [$\Delta(1940)D_{33}$],
$\Delta(1930)D_{35}$ \\
$^{2}10_J[70,2^+]$ & (0,0) & 1945 & [$\Delta(2000)F_{35}$] \\
$^{4}10_J[56,0^+]$ & (1,0) & 1646 & $\Delta(1600)P_{33}$ \\
$^{2}10_J[70,1^-]$ & (1,0) & 1977 & $\Delta(1900)S_{31}$ \\
$^{2}10_J[70,0^+]$ & (0,1) & 1786 & [$\Delta(1750)P_{31}$] \\
& & & \\
\hline
\end{tabular}
\end{table}

\clearpage

\clearpage

\clearpage

\clearpage

\clearpage

\clearpage

\begin{figure}
\centering
\setlength{\unitlength}{1.0pt}
\begin{picture}(350,200)(0,0)
\thicklines
\put ( 75, 50) {\line ( 1,0){200}}
\put ( 75, 50) {\line ( 1,2){100}}
\put (275, 50) {\line (-1,2){100}}
\put ( 50, 30) {Anharmonic}
\put ( 50, 15) {vibrator ($\beta=0$)}
\put (250, 30) {$\gamma$ unstable}
\put (250, 15) {rotor ($\beta>0$)}
\put ( 90,270) {Axial rotor  ($\beta>0$, $\gamma=0$ or $\pi/3$)}
\end{picture}
\vspace{15pt}
\caption[Symmetry triangle]{Phase triangle of the IBM.}
\label{shape1}
\end{figure}

\clearpage

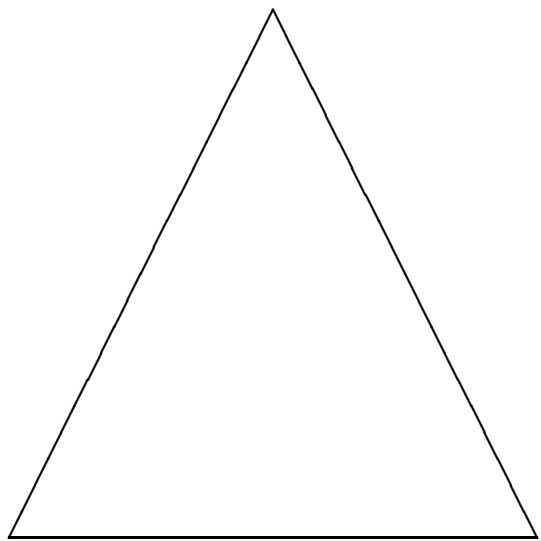
\begin{figure}
\centering
\setlength{\unitlength}{1.0pt}
\begin{picture}(240,180)(0,0)
\thinlines
\put (  0,  0) {\line(1,0){240}}
\put (  0,180) {\line(1,0){240}}
\put (  0,  0) {\line(0,1){180}}
\put (240,  0) {\line(0,1){180}}
\thicklines
\put ( 70, 30) {\line(1,0){20}}
\put ( 70, 85) {\line(1,0){20}}
\put ( 70,150) {\line(1,0){20}}
\thinlines
\put ( 30, 25) {$n=0$}
\put ( 30, 80) {$n=1$}
\put ( 30,145) {$n=2$}
\put ( 95, 25) {$0^+_S$}
\put ( 95, 80) {$1^-_M$}
\put ( 95,145) {$2^+_S~, \, 2^+_M ~, \, 1^+_A ~, \, 0^+_S ~, \, 0^+_M$}
\put (180, 25) {$U(6)$}
\end{picture}
\vspace{15pt}
\caption[Anharmonic oscillator]
{Schematic spectrum with $U(6)$ symmetry. 
The masses are calculated using Eq.~(\ref{mass1}) 
with $\epsilon_1>0$, $\epsilon_2>0$ and 
$\alpha=\kappa=\kappa^{\prime}=0$. The number of bosons is $N=2$.}
\label{anhosc}
\end{figure}
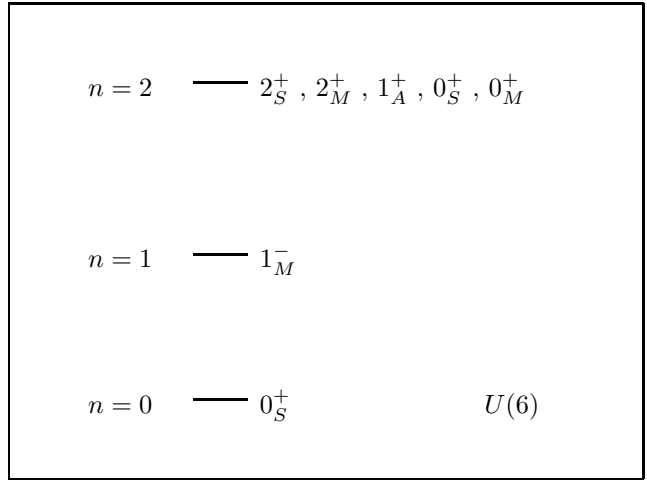

\clearpage

\clearpage

\clearpage

\clearpage

\begin{figure}
\centering
\setlength{\unitlength}{1.0pt}
\begin{picture}(350,200)(0,0)
\thicklines
\put ( 75, 50) {\line ( 1,0){200}}
\put ( 75, 50) {\line ( 1,2){100}}
\put (275, 50) {\line (-1,2){100}}
\put ( 50, 30) {Anharmonic}
\put ( 50, 14) {vibrator ($r=0$)}
\put (225, 30) {Deformed oscillator ($r>0$)}
\put (225, 15) {(or $\chi$, $\theta$ unstable rotor)}
\put ( 75,270) {Oblate top ($r>0$, $\chi=\pi/4$, $\theta=\pi/2$)}
\end{picture}
\vspace{15pt}
\caption[Shape triangle]
{Phase triangle of $U(7)$ with $S_3$ invariance.}
\label{shape2}
\end{figure}

\end{document}